\documentclass[twocolumn,showpacs,showkeys,preprintnumbers,pra,smaller,
superscriptaddress]{revtex4}
\usepackage{amssymb,amsmath,graphicx,dcolumn,bm,epsfig,float}
\usepackage[utf8]{inputenc}
\usepackage[T1]{fontenc}
\usepackage[colorlinks,linkcolor=black,citecolor=blue]{hyperref}

\newcommand{\opn}[1]{\ensuremath{\operatorname{#1}}}

\begin{document}

\preprint{Physical Review A}
\title{Distant entanglement enhanced in $\mathcal{PT}$-symmetric optomechanics}
\author{C. Tchodimou}
\email{c.tchodimou@univ-ndere.cm}
\affiliation{Laboratory of Photonics, Department of Physics, Faculty of Science,
University of Ngaoundere , P.O. Box 454, Ngaoundere , Cameroon}
\author{P. Djorwe}
\email{philippe.djorwe@univ-ndere.cm,\\djorwepp@gmail.com}
\affiliation{Laboratory of Photonics, Department of Physics, Faculty of Science,
University of Ngaoundere , P.O. Box 454, Ngaoundere , Cameroon}
\affiliation{Laboratory of Modelling and Simulation in Engineering, Biomimetics and
Prototypes, Department of Physics, Faculty of Science,\\
University of Yaounde I, P.O. Box 812, Yaounde, Cameroon}
\author{S.G. Nana Engo}
\email{snana@univ-ndere.cm}
\affiliation{Laboratory of Photonics, Department of Physics, Faculty of Science,
University of Ngaoundere , P.O. Box 454, Ngaoundere , Cameroon}

\begin{abstract}
We study steady-state continuous variable entanglement in a three-mode 
optomechanical system consisting of an active optical cavity (gain) coupled to a 
passive optical cavity (loss) supporting a mechanical mode. For a driving laser 
which is blue-detuned, we show that coupling between optical and mechanical 
modes is enhanced in the unbroken-$\mathcal{PT}$-symmetry regime. We analyze the 
stability and this shows  that steady-state solutions are more stable in the 
gain and loss systems.  We use these stable solutions to generate distant 
entanglement between the mechanical mode and the optical field inside the gain 
cavity. It results in a giant enhancement of entanglement compared to what is 
generated in the single lossy cavity. This work offers the prospect of exploring 
quantum state engineering and quantum information in such systems. Furthermore, 
such entanglement opens up an interesting possibility to study spatially 
separated quantum objects.
\end{abstract}

\pacs{ 42.50.Wk, 42.50.Lc, 42.50.Pq}
\keywords{Optomechanics, $\mathcal{PT}$-symmetry, entanglement}
\maketitle

\date{\today}


\section{Introduction}

Significant advances in the study of light-matter interaction have been carried 
out through optomechanics \cite{[1]}.  Ground-state cooling of macroscopic 
objects \cite{[2],[3],[4]}, squeezing quantum noises below the quantum standard 
limit \cite{[5],[6]}, quantum entanglement \cite{[7]}-\cite{[21b]}, and 
macroscopic quantum superposition \cite{[22]} have been deeply improved. This 
promotes a wide variety of quantum applications, such as quantum Sensors 
\cite{[23]}, quantum informations processing \cite{[24]}, quantum metrology 
\cite{[25]} and quantum computational tasks \cite{[26]}. However, there are 
still some limitations to fully handle certain quantum  optomechanical 
applications and so numerous efforts are ongoing. Indeed, quantum entanglement 
is often limited by various factors such as the stability conditions that place 
constraints on the magnitude of the effective optomechanical couplings 
\cite{[27]}-\cite{[29]} and the amplification effect in the unstable regime 
\cite{[30]}. In particular, thermal noise of the mechanical modes can strongly 
impair the generation of such nonclassical states.

Very recently, systems described by non-Hermitian Hamiltonians (see \cite{[31]} 
and the references therein), have been used to engineer cavity optomechanics 
(COM) \cite{[32]}-\cite{[36]}. This has led to low-power phonon emissions 
\cite{[32]}, emergency of chaos at low-power  threshold \cite{[33]} and 
non-reciprocal topological energy transfer \cite{[35]}. These three-mode COM 
systems consist of an active optical cavity (gain) coupled to a passive optical 
cavity (loss) supporting a mechanical mode \cite{[37]}-\cite{[39]}. Taking 
advantage of the intriguing properties of these systems, we aim to improve the 
stability  and to enhance the magnitude of the effective optomechanical 
couplings. Two regimes can be identified, i.e., the unbroken-$ 
\mathcal{PT}$-symmetry regime which happens for strong optical tunneling rate 
and the broken-$\mathcal{PT}$-symmetry regime which happens for weak optical 
tunneling rate \cite{[40]}. The transition between these two regimes corresponds 
to the exceptional point (EP). It has been shown that the intracavity photon 
number is significantly improved in such systems even at low driving power 
\cite{[32]}. This leads to an enhancement of the effective optomechanical 
couplings. That yields a robust entanglement generation between the mechanical 
mode and the optical field inside the gain cavity. Furthermore, this 
entanglement is improved by adding a parametric amplifier (PA) in the loss 
cavity. It should be noted that the use of a PA for entanglement generation has 
been considered in \cite{[14],[14a]}. 

Our findings can be stated as follows. For a driving laser which is 
blue-detuned, the effective optomechanical coupling is enhanced and is more 
stable in the unbroken-$\mathcal{PT}$-symmetry regime. It results in a strong 
entanglement between the mechanical mode and the optical field inside the gain 
cavity. Such quantum correlation between distant modes is known as distant 
entanglement \cite{[41],[42]}, and might present an interesting possibility to 
study spatially separated quantum objects. Our results are different from those 
in \cite{[41],[42]} where the cavities that are used are lossy cavities, and 
therefore do not exhibit  $\mathcal{PT}$-symmetry. Owing to the presence of 
$\mathcal{PT}$-symmetry in our configuration, the generated entanglement is 
enhanced compared to what is done in \cite{[41],[42]}. Furthermore, the addition 
of a squeezing element in the loss cavity improves the magnitude of the 
effective optomechanical coupling that induces some amount of entanglement as 
well.

The work is organized as follows. In Sec. \ref{sec:Dyn}, the system and the 
dynamical equations  are described. The stability analysis is presented in Sec. 
\ref{sec:Stab}. Section \ref{sec:Entang} is devoted to the continuous variable 
(CV) entanglement generation and their robustness against the thermal 
decoherence. We conclude the work in Sec. \ref{sec:Concl}.

\section{System and dynamical equations}

\label{sec:Dyn}

\begin{figure*}[tbh]
\centering
\par
\begin{center}
\resizebox{0.6\textwidth}{!}{
\includegraphics{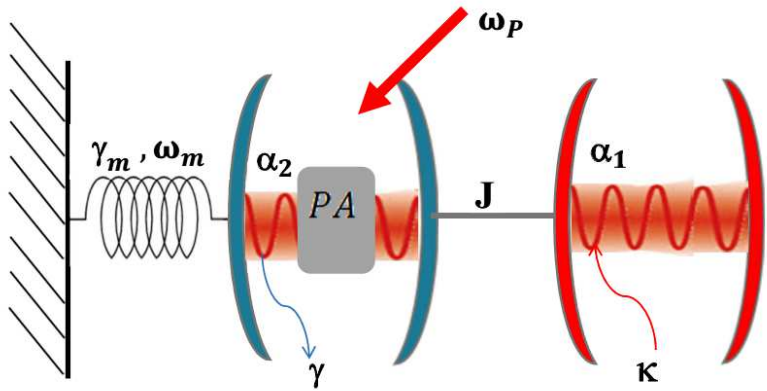}} 
\end{center}
\caption{Setup of the gain and loss COM used. An active optical cavity (gain) is 
coupled to a passive optical cavity (loss) supporting a mechanical mode. A 
squeezing element (PA) is inserted inside the loss cavity.}
\label{fig:Figa}
\end{figure*}

We consider a system of two coupled microresonators (see Fig. \ref{fig:Figa}), 
one with an optical gain $\kappa$ (active optical cavity) and the other with 
loss $\gamma$ (passive optical cavity) \cite{[37]}-\cite{[39]}. In such system, 
both the coupling strength $J$ and the gain-to-loss ratio of the resonators can 
be tuned, as experimentally demonstrated in \cite{[39]}. The mechanical mode, 
with frequency $\omega_m$ and effective mass $m$, contained in the passive 
resonator is driven by an external field having a frequency $\omega_p$. The 
Hamiltonian of this system can be written as ($\hbar =1$) \cite{[31]},

\begin{equation}
\begin{cases}
H=H_0+H_{int}, \\ 
H_0=\omega_m\beta^\dag\beta -\Delta (\alpha_1^\dag\alpha_1 
+\alpha_2^\dag\alpha_2), \\ 
H_{int}=J(\alpha_1^\dag\alpha_2+\alpha_2^\dag\alpha_1)-g\alpha_2^\dag\alpha_2 
(\beta ^\dag+\beta)+\sqrt{\gamma}\epsilon^{in}(\alpha_2^\dag+\alpha_2).
\end{cases}
\label{1}
\end{equation}
The lowering operators $\beta$, $\alpha_1$ and $\alpha_2$ describe the 
mechanical resonator, the active optical cavity and the passive optical cavity 
respectively. The COM coupling coefficient is given by $g$. We choose a driving 
pump whose frequency is blue-detuned $\Delta =\omega_p -\omega_c>0$, where 
$\omega_c$ is the cavity frequency. The Hamiltonians representing the optical 
gain and loss and the mechanical damping are not explicitly shown here.

From the above Hamiltonian, one derives the following set of dynamical 
equations, 
\begin{equation}
\begin{cases}
\dot{\alpha}_1=\left( i\Delta +\frac{\kappa }{2}\right) \alpha_1-iJ\alpha_2 
+\sqrt{\kappa }\alpha_1^{in}, \\ 
\dot{\alpha}_2=\left[ i(\Delta+g(\beta^\dag+\beta))-\frac{\gamma}{2}\right] 
\alpha_2-iJ\alpha_1-i\sqrt{\gamma }\epsilon ^{in}, \\ 
\dot{\beta}=-(i\omega_m+\frac{\gamma_m}{2})\beta +ig\alpha_2^\dag\alpha_2 
+\sqrt{\gamma_m}\beta ^{in},
\end{cases}
\label{2}
\end{equation}
where $\gamma_m$ is the mechanical damping and $\beta ^{in}$ stands for
the thermal driving at finite environmental temperature $T$. The driving
field $\epsilon ^{in}=\alpha ^{in}+\alpha_2^{in}$ consists of a coherent
amplitude $\alpha ^{in}$ and a vacuum noise operator $\alpha_2^{in}$.
Similarly, the vacuum noise operator associated to the field $\alpha_1$
is $\alpha_1^{in}$.

In order to gain an insight into the behaviors of the system that we are
interested in, namely the unbroken-$\mathcal{PT}$-symmetry and the broken-$%
\mathcal{PT}$-symmetry regimes, we consider only the optical modes in Eq. (%
\ref{2}) and ignore the driving \cite{[37]}. By diagonalizing these optical
modes, we obtain the eigenfrequencies of the two supermodes as well as the
associated linewidths as given by the real and imaginary parts,
respectively, of the complex frequencies

\begin{equation}
\omega _{\pm }=\frac{1}{4}\left(4i\Delta-(\gamma-\kappa)\pm 
\sqrt{(\gamma +\kappa)^2-16J^2}\right).  \label{3}
\end{equation}

For a strong optical tunneling rate, i.e., $J>(\gamma+\kappa)/4$, the system 
exhibits the purely optical unbroken-$ \mathcal{PT}$-symmetry regime while the 
broken-$\mathcal{PT}$-symmetry regime holds for weak optical tunneling rate 
with $J<(\gamma+\kappa)/4$ \cite{[32],[40]}. The transition between these 
two regimes, i.e., $J=(\gamma+\kappa)/4$ \cite{[32]}, corresponds to the 
exceptional points where the two eigenfrequencies coalesce.

As the aim is to enhance entanglement, we add an additional parametric amplifier 
(PA) in the loss cavity as indicated in Fig. \ref{fig:Figa}. This squeezing 
element is described by the Hamiltonian, \begin{equation} H_\chi=i\chi \left( 
e^{i\theta}(\alpha_2^\dag)^2-e^{-i\theta }(\alpha_2)^2 \right), \end{equation} 
where $\chi$ is the gain of the PA while $\theta$ is the phase of the pump 
driving it. We have set $\theta=0$ in the whole work while the gain $\chi$ can 
be tuned.

In the next section, we study the stability of the steady-state solutions
in order to quantify distant entanglement.

\section{Steady states and stability analysis}

\label{sec:Stab}
 
For $|\langle \alpha_2\rangle| \gg 1$, the operators in Eq. (\ref{2}) can be 
expanded as their mean values plus a small amount of fluctuations. This yields 
the steady states of our system, by linearizing the field operators around their 
steady-state values $\mathcal{O}(t)=O_s+\delta \mathcal{O}(t)$, where 
$\mathcal{O}\equiv (\alpha_1$, $\alpha_2$, $\beta)$.  The stability analysis of 
these steady-state solutions should be addressed, since the important feature of 
these solutions is their stability. Indeed, any steady-state solution is 
dynamically meaningless unless it is stable. We study this stability through 
linear stability analysis \cite{[43]}. Throughout the work, we use the following 
experimentally achievable parameters \cite{[37]}-\cite{[39]}, $\omega_m/ 2\pi 
=23\opn{MHz}$, $\gamma/2\pi =1\opn{MHz}$, $g=7.4\times10^{-5}\gamma $ and 
$\gamma_m=1.63\times 10^{-3}\gamma$. The parameters $\kappa$ and $J$ will be 
tuned according to \cite{[39]}. From the linearization, the dynamical 
fluctuation of the system can be described by the compact equation, 
\begin{equation}
 \dot{v}(t)=M v(t)+z(t), \label{5}
\end{equation}
where $v(t)$ is the vector of the fluctuations,  
$v(t)=\left(\delta\beta(t),\delta
\beta^\dag(t),\delta \alpha_1(t),\delta \alpha_1^\dag(t),\delta\alpha_2(t), 
\delta \alpha_2^\dag(t)\right)^T$, and its associated noise vector is
\begin{widetext}
\begin{equation}
z(t)=\left( \sqrt{\gamma_m}\delta \beta^{in}(t),\sqrt{\gamma_m}\delta \beta 
^{in\dag }(t),\sqrt{\kappa }\delta \alpha_1^{in}(t),\sqrt{\kappa }\delta 
\alpha_1^{in\dag }(t),-\sqrt{\gamma }\delta \alpha_2^{in}(t),\sqrt{\gamma 
}\delta \alpha_2^{in\dag}(t)\right)^T. 
\end{equation}
\end{widetext}
The matrix $M$ stands for the Jacobian of the system and is given by,
\begin{equation}
M=\begin{bmatrix}
-\Omega_\beta & 0 & 0 & 0 & iG_2 & iG_2 \\ 
0 & -\Omega_\beta^\ast & 0 & 0 & -iG_2 & -iG_2 \\ 
0 & 0 & \Omega_{\alpha_1} & 0 & -iJ & 0 \\ 
0 & 0 & 0 & \Omega _{\alpha_1}^{\ast } & 0 & iJ \\ 
iG_2 & iG_2 & -iJ & 0 & \Omega_{\alpha_2} & 2\chi e^{i\theta } \\ 
-iG_2 & -iG_2 & 0 & iJ & 2\chi e^{-i\theta } & \Omega_{\alpha_2}^\ast
\end{bmatrix}
,
\end{equation}
where $G_2=g|\alpha_{2,s}|$ is the direct effective optomechanical coupling 
strength, and we have defined $\Omega_\beta =i\omega_m+\frac{\gamma_m}{2}$, 
$\Omega_{\alpha_1}=i\Delta +\frac{\kappa }{2}$, and $\Omega_{\alpha_2} 
=i\tilde{\Delta}-\frac{\gamma }{2}$. The stability analysis of the system can 
be done based on the eigenvalues of the matrix $M$. In fact, a given 
steady-state solution is stable if all eigenvalues of $M$ have a negative real 
part. Otherwise, the steady-state will not converge towards a fixed point, and 
might exhibit a limit cycle or chaotic behavior \cite{[33]}. These cases are 
not considered in our analysis. The frequency shift $\delta\Delta 
=2g\Re(\beta_s)$ induces the nonlinear detuning $\tilde{\Delta}=\Delta 
+\delta\Delta$.

\begin{figure*}[tbh]
\centering
\par
\begin{center}
\resizebox{0.38\textwidth}{!}{
\includegraphics{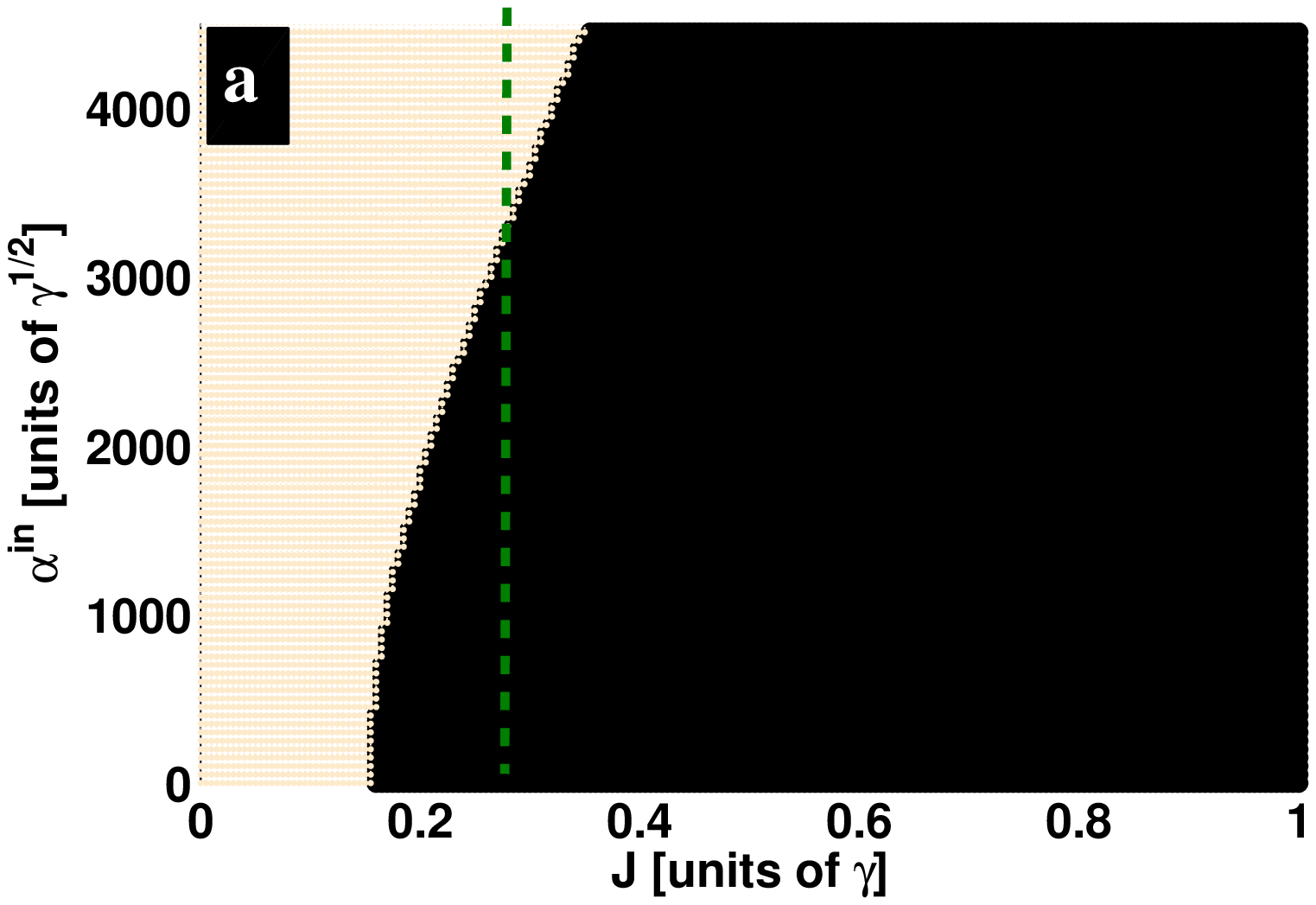}} 
\resizebox{0.38\textwidth}{!}{
\includegraphics{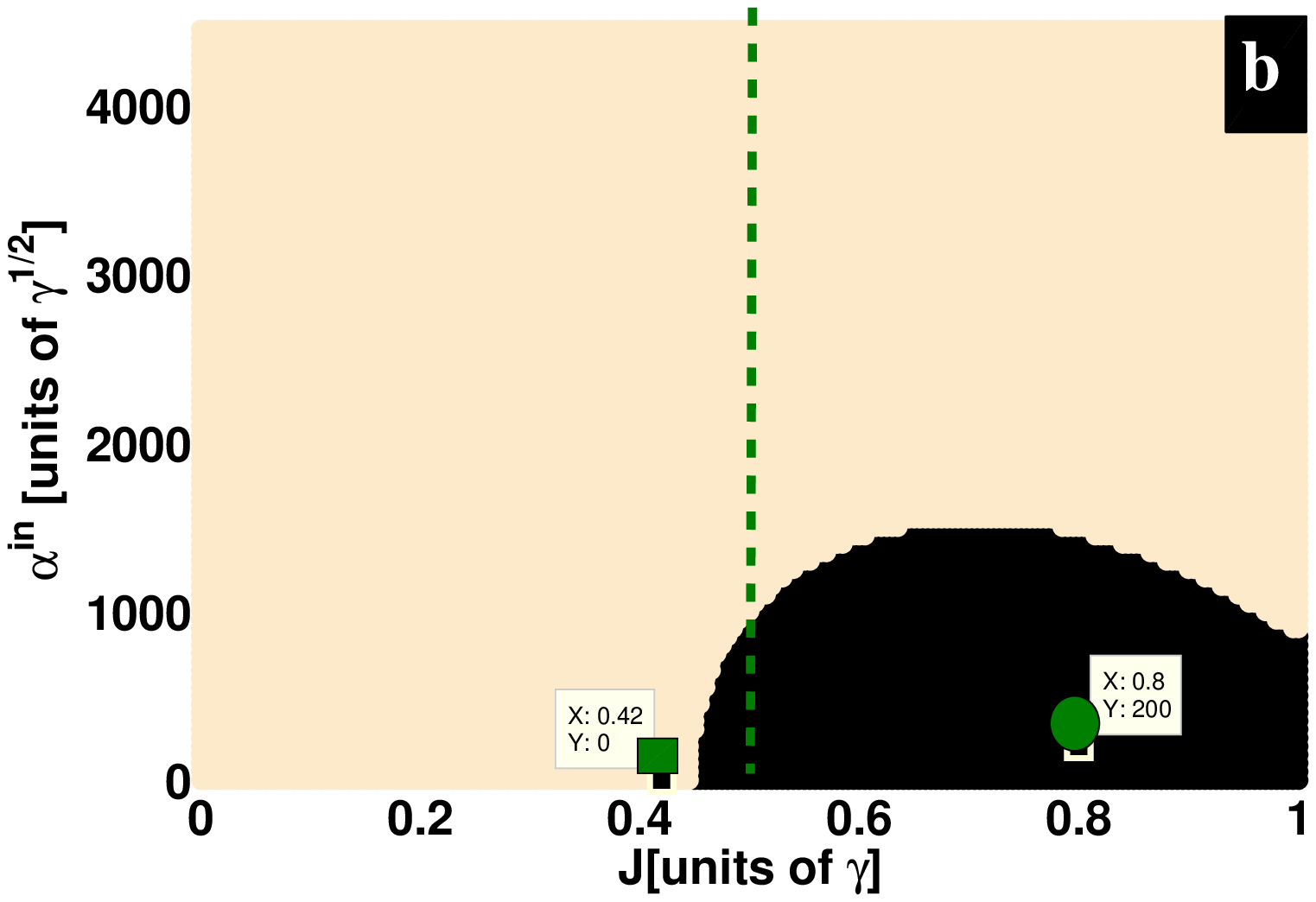}} 
\resizebox{0.38\textwidth}{!}{
\includegraphics{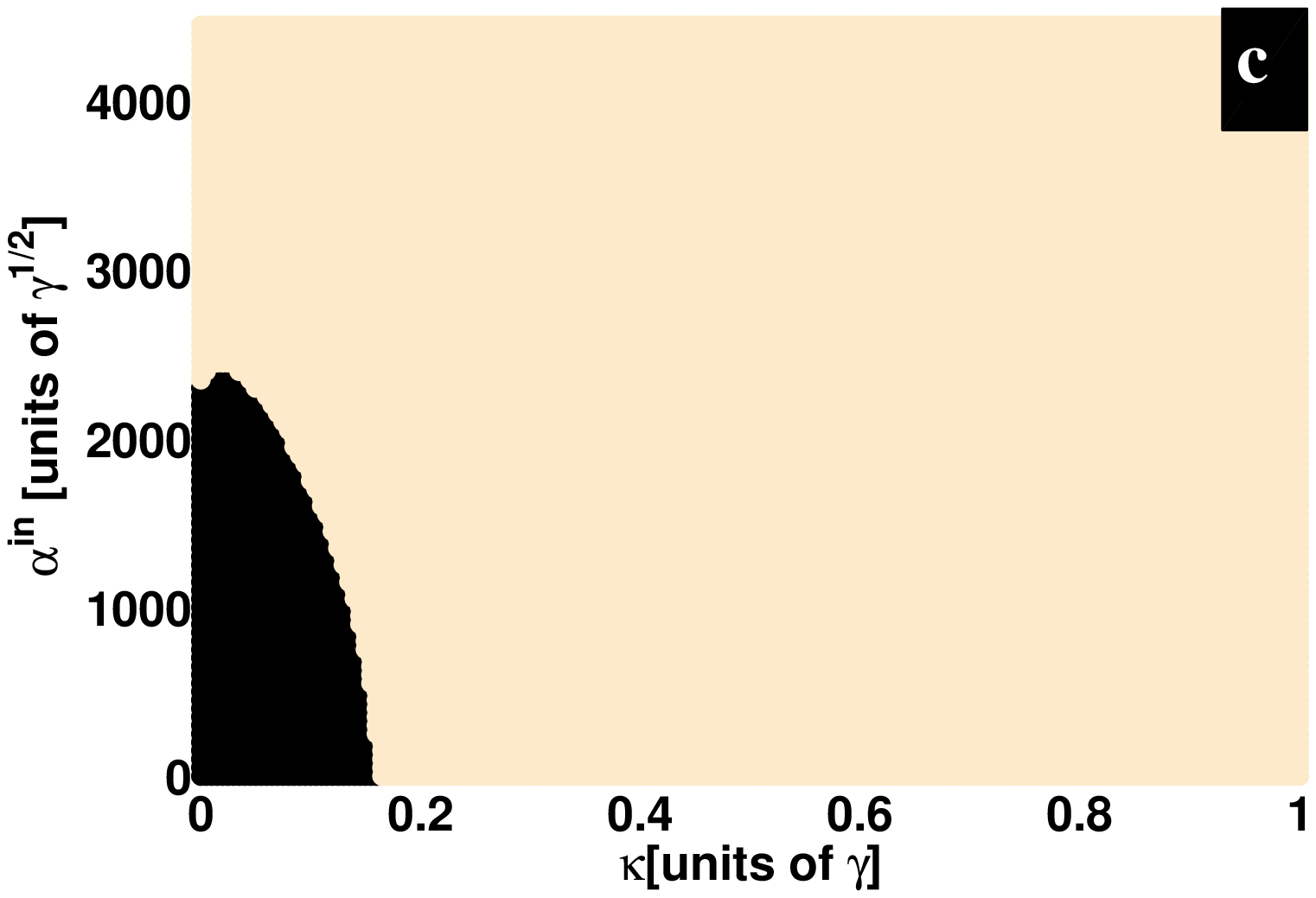}} 
\resizebox{0.38\textwidth}{!}{
\includegraphics{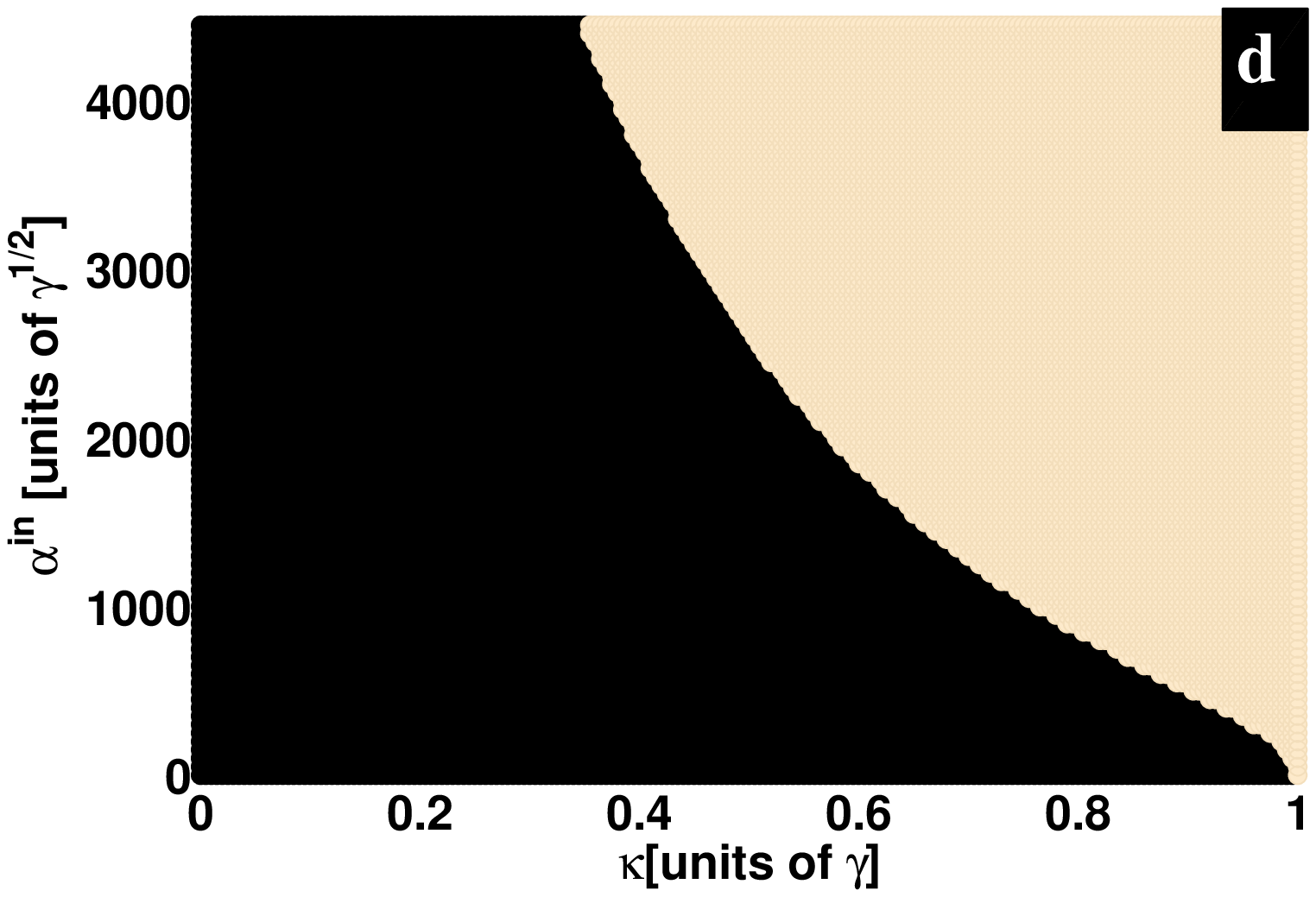}}
\end{center}
\caption{Basins of stability for the steady states. (a),(b). The evolution of 
stable (dark area) and unstable (light area) zones as the system gets closer to 
the balanced gain and loss limit. (a) $\kappa =0.1\gamma$; (b) $\kappa 
=0.8\gamma$. (c),(d) The evolution of stable (dark area) and unstable (light 
area) zones as the tunneling coupling $J$ increases. (c) $J=0.2\gamma$; (d) 
$J=\gamma$. These figures are obtained for $\chi =0$.}
\label{fig:Fig1}
\end{figure*}

\begin{figure*}[tbh]
\centering
\par
\begin{center}
\resizebox{0.38\textwidth}{!}{
\includegraphics{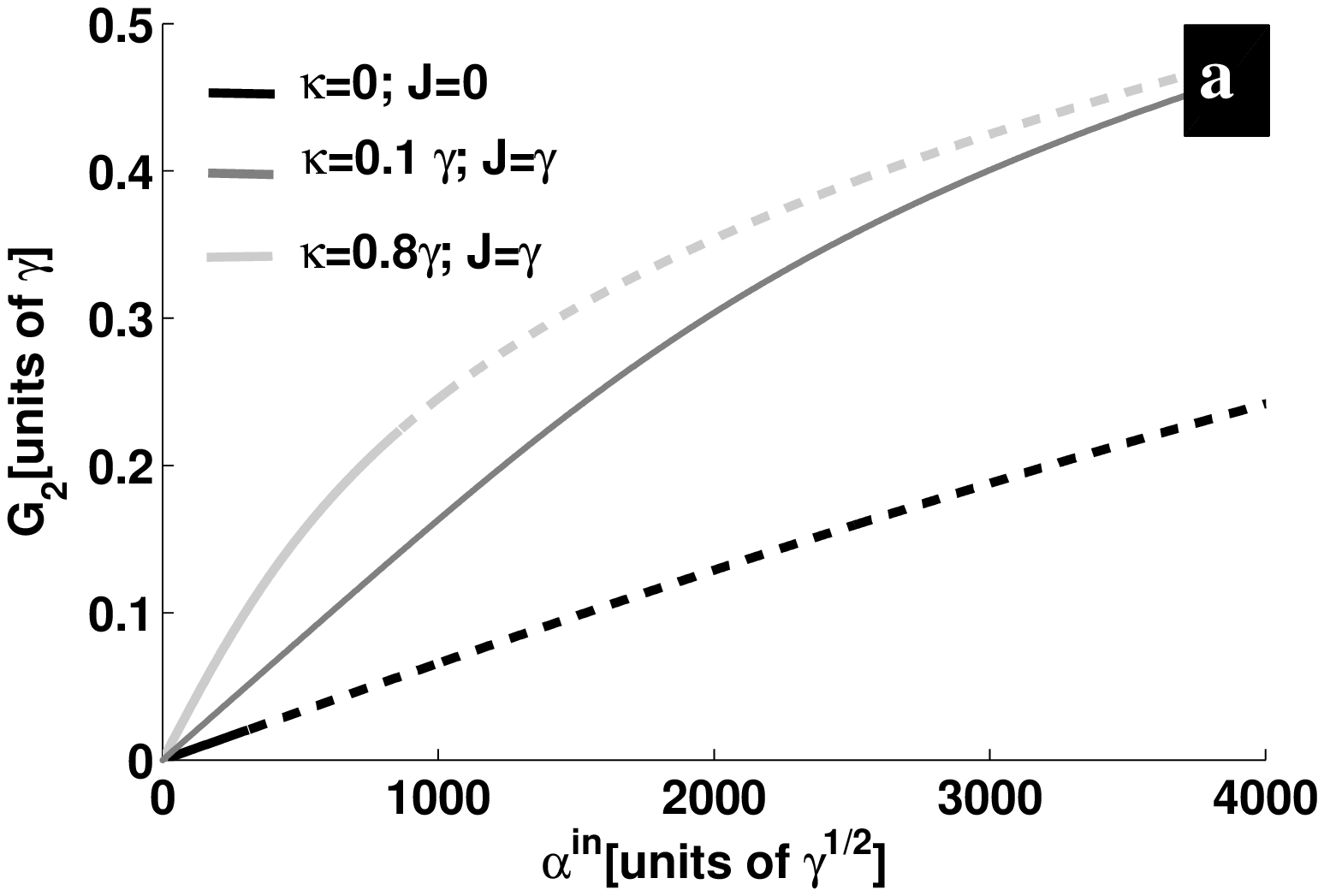}} 
\resizebox{0.38\textwidth}{!}{
\includegraphics{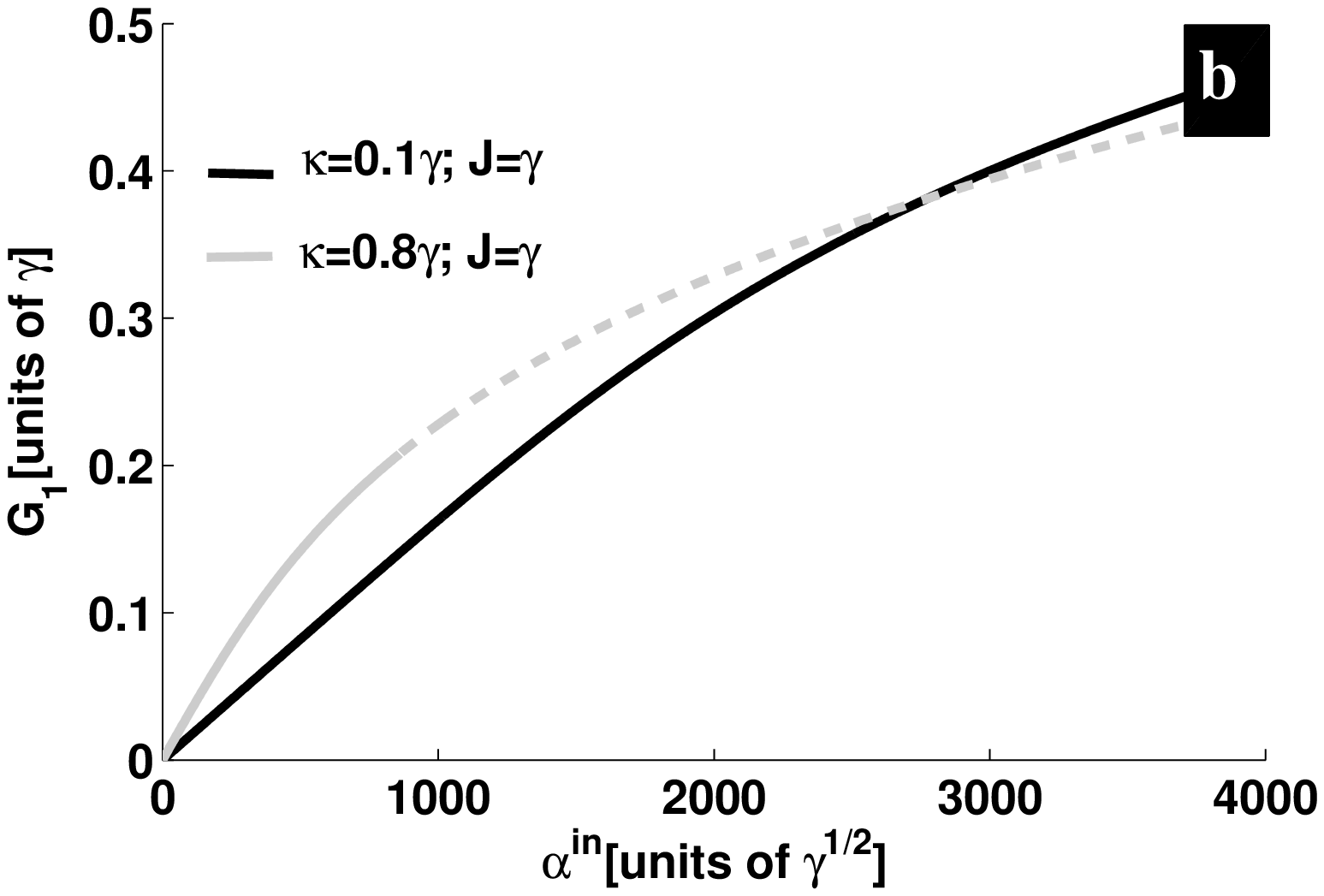}} 
\resizebox{0.38\textwidth}{!}{
\includegraphics{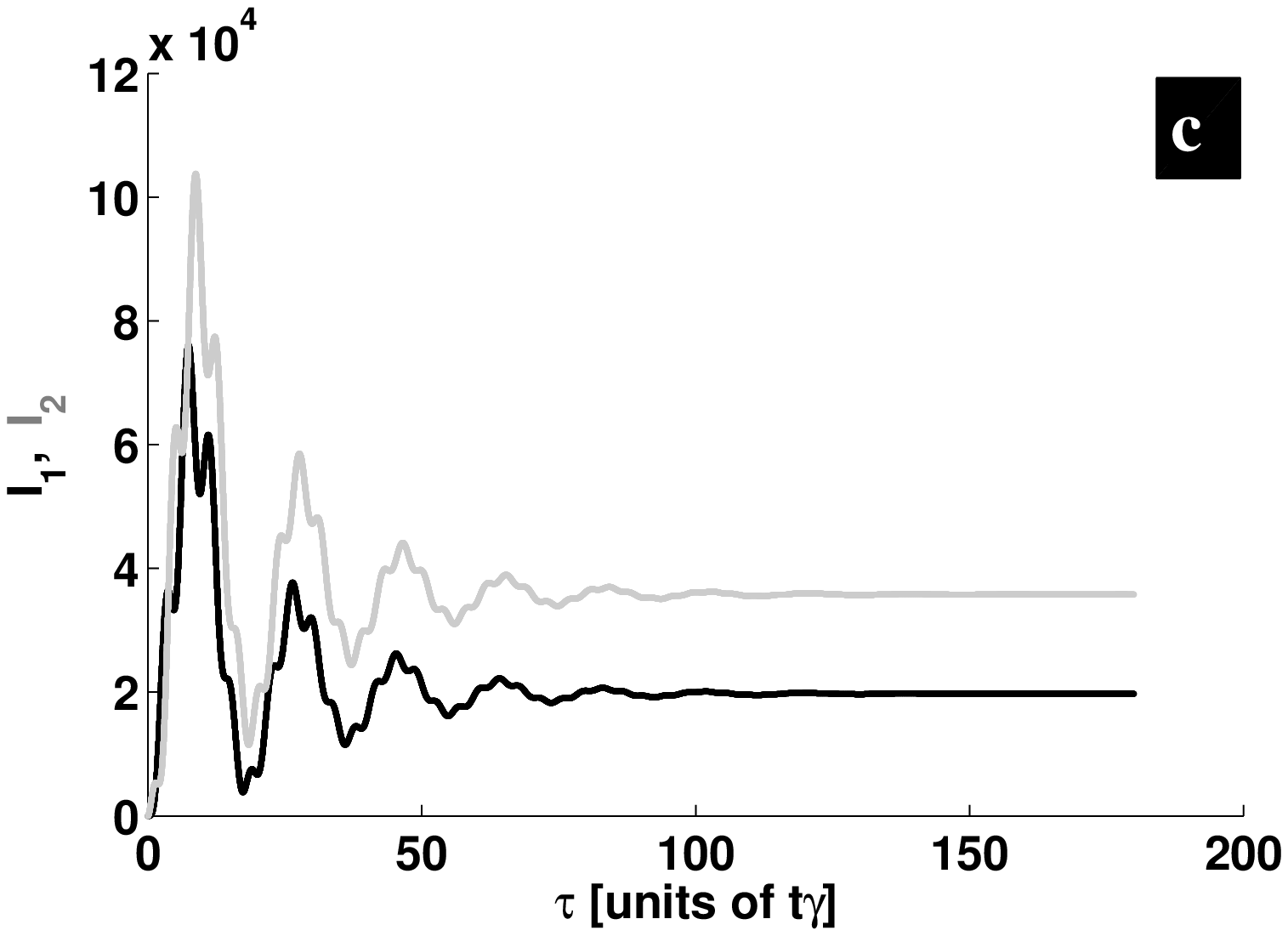}} 
\resizebox{0.38\textwidth}{!}{
\includegraphics{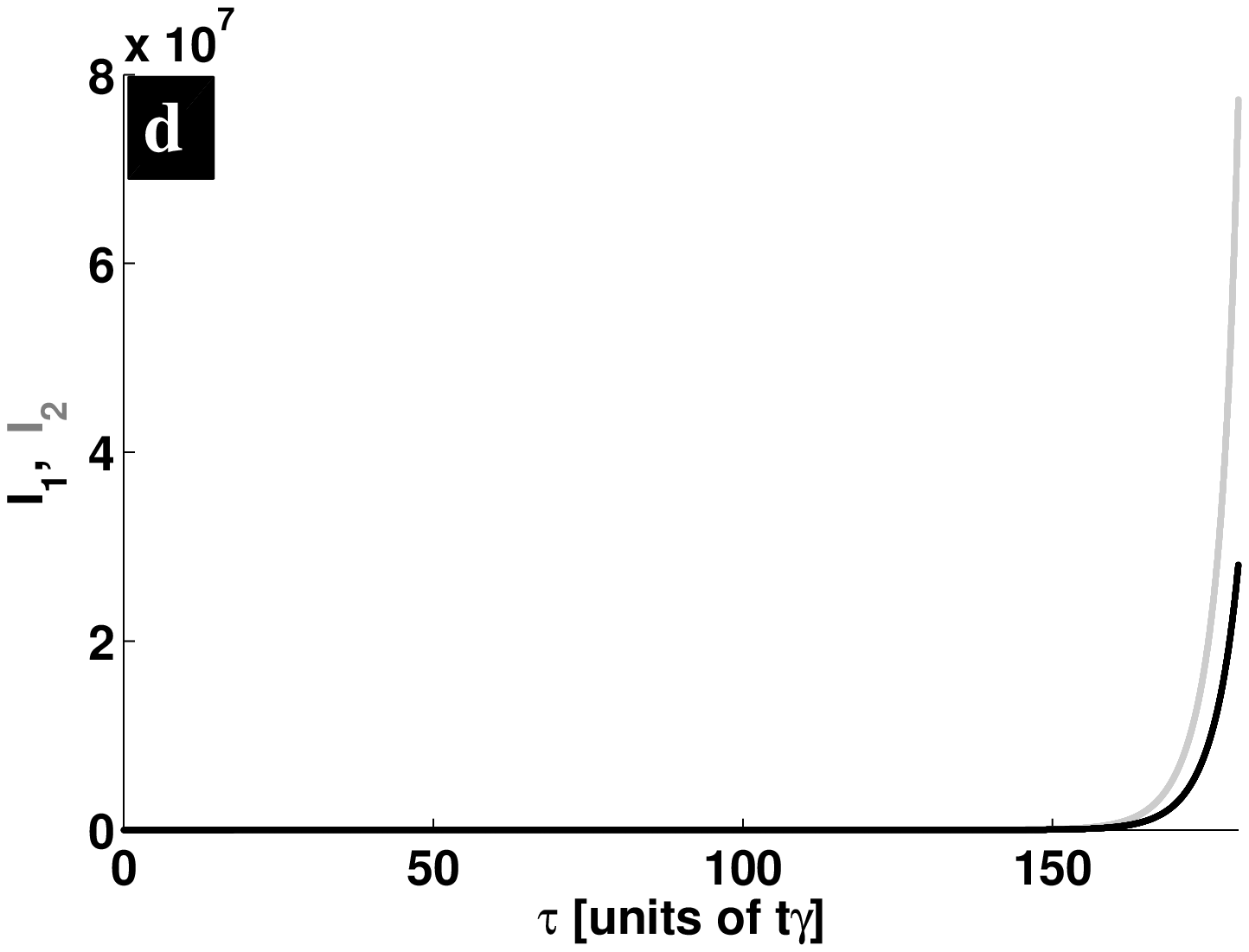}}
\end{center}
\caption{Evolution of steady states accordingly to the stability shown in 
Fig. 
\ref{fig:Fig1}. (a),(b) $G_i=g|\alpha_{i,s}|,\, i=1,2$, are the couplings 
strength. Solid lines are stable while dashed line are unstable. (c),(d) Time 
evolution of steady states having stable and unstable dynamics respectively, 
with $I_i=|\alpha^*_i\alpha_i|,\,i=1,2$.  The parameters are (c) 
$\alpha^{in}=10^2 \sqrt{\gamma }$, $J=0.8\gamma $ and $\kappa =0.8\gamma$, and 
(d) $\alpha^{in}=10^{-5}\sqrt{\gamma }$, $J=0.42\gamma$ and $\kappa =0.8\gamma 
$. For all of these figures, $\chi =0$ and the index $i=1$ is related to the 
lossy cavity whereas the index $i=2$ stands for the gain cavity.}
\label{fig:Fig2}
\end{figure*}

The linear stability of our system is mapped out through the basins of stability 
shown in Fig. \ref{fig:Fig1}. These figures are obtained in the absence of the 
PA ($\chi =0$). The steady states are stable for the range of parameters located 
in the blue space, while the red area corresponds to the parameters leading to 
unstable fixed points. It follows that the unstable area widens as the system 
moves towards the balanced gain and loss [compare Figs. \ref{fig:Fig1}(a) and 
\ref{fig:Fig1}(b)]. We also remark that the stability of the system is extended 
to large driving strength compared to the single lossy cavity (compare black and 
other colors in Fig. \ref{fig:Fig2}a). From Fig. \ref{fig:Fig1}c and Fig. 
\ref{fig:Fig1}d, it appears that the stability of the system improves as the 
tunneling coupling $J$ increases. Still, the stable steady states are shifted 
towards relatively large  driving strength. In light of this discussion on the 
stability in active-passive COM, we conclude that (i) the steady-state solutions 
are stable for the imbalanced gain and loss system (weak $\kappa$), and (ii) 
stability can be improved by increasing the tunneling rate $J$, by pushing the 
system in the unbroken-$\mathcal{PT}$-symmetry regime. Another observed feature 
is the stability of the fixed points related to the EP. Such points are located 
along the green dashed lines in Figs. \ref{fig:Fig1}(a) and \ref{fig:Fig1}(b), 
for $\kappa =0.1\gamma$ (or $J=0.275\gamma$) and $\kappa =0.8\gamma$ (or 
$J=0.45\gamma$) respectively. It can be seen that the solutions along the EP 
lose their stability as the system approaches the gain-loss balance. Indeed, we 
have observed that at the exact gain-loss balance, the EP is completely in the 
unstable zone (no represented). From this analysis and in light of Figs. 
\ref{fig:Fig1}(a) and \ref{fig:Fig1}(b), it is shown that the stable steady 
states in active-passive COM are mostly located in the 
unbroken-$\mathcal{PT}$-symmetry regime.

As the gain cavity is coupled to the mechanical resonator through the loss 
cavity, one can define the resulting (distant) coupling as $G_1=g|\alpha 
_{1,s}|$. This suggests the idea of investigating distant entanglement between 
the gain cavity and the mechanical resonator. Figures \ref{fig:Fig2}(a) and 
\ref{fig:Fig2}(b) show the couplings $G_2$ and  $G_1$ versus the driving pump 
$\alpha ^{in}$, respectively. Full lines are stable while dashed lines are 
unstable, in accordance with the stability shown in  Fig. \ref{fig:Fig1}. For 
$\chi =0$, Figs. \ref{fig:Fig2}(a) and \ref{fig:Fig2}(b) show the 
enhancement of the couplings $G_2$ and  $G_1$ as well as the improvement of the 
system's stability (gray colors) compared to the conventional COM case [black 
color in Fig. \ref{fig:Fig2}(a)]. As a result, the steady states are more 
stable in the systems having gain and loss, and this paves a way to use such 
systems to enhance the quantum effect as for entanglement here. As the system 
moves from imbalanced gain and loss to the balanced case, the magnitude of the 
couplings $G_{1,2}$ slightly increases while the stability is impaired [see 
Figs. \ref{fig:Fig2}(a) and  \ref{fig:Fig2}(b)]. We have remarked that $\chi$ 
slightly enhances the couplings $G_{1,2}$ (not represented), and this effect 
will be highlighted through the entanglement later on. It is shown that, (i) 
stability is improved in the unbroken-$\mathcal{PT}$-symmetry regime and (ii) 
the couplings $G_{1,2}$ are enhanced when approaching gain and loss balance. 
Based on this discussion, we have determined which regime to study distant 
entanglement. But first, we have performed the validity of the above stability 
analysis through a numerical simulation of Eq. (\ref{2}). Figure 
\ref{fig:Fig2}(c) corresponds to the stable steady state solution indicated by 
the green dot in Fig. \ref{fig:Fig1}(b). Figure \ref{fig:Fig2}(d) shows the 
dynamical state of the unstable fixed point localized in the vicinity of the 
green square in Fig. \ref{fig:Fig1}(b). We can see that the solution is 
attracted to a fixed point in Fig. \ref{fig:Fig2}(c) while it grows 
exponentially in Fig. \ref{fig:Fig2}(d). This exponential growth is reminiscent 
of instabilities, and the long-time study of such behavior yields a limit cycle 
or chaotic dynamics. This numerical investigation ensures the veracity of our 
stability analysis.

\section{Entanglement generation}

\label{sec:Entang}

To measure the CV entanglement between the gain cavity and the mechanical 
modes, we use the standard ensemble method, which consists of computing the 
logarithmic negativity through the quantum fluctuations of the system's 
quadratures \cite{[7],[8]}. From the set of fluctuation equations given in Eq. 
(\ref{5}), we can define the vector of quadratures $u(t)=\left(\delta 
x(t),\delta p(t),\delta I_1(t),\delta \varphi_1(t),\delta I_2(t),\delta 
\varphi_2(t)\right)^T$ and the vectors of noises $ n(t)=\left( \delta 
I_{x}^{in}(t),\delta I_p^{in}(t), \delta I_1^{in}(t),\delta 
\varphi_1^{in}(t),\delta I_2^{in}(t),\delta \varphi_2^{in}(t)\right)^T$. Thus, 
the system's quadratures can be written in the compact form
\begin{equation}
\dot{u}(t)=Au(t)+n(t), \label{10}
\end{equation}%
with the correlation matrix 
\begin{equation}
A=\begin{bmatrix}
-\frac{\gamma_m}{2} & \omega_m & 0 & 0 & 0 & 0 \\ 
-\omega_m & -\frac{\gamma_m}{2} & 0 & 0 & 2G_2 & 0 \\ 
0 & 0 & \frac{\kappa }{2} & -\Delta  & 0 & J \\ 
0 & 0 & \Delta  & \frac{\kappa}{2} & -J & 0 \\ 
0 & 0 & 0 & J & (2\chi \cos\theta-\frac{\gamma }{2}) & (2\chi \sin\theta 
-\tilde\Delta)  \\ 
2G_2 & 0 & -J & 0 & (2\chi \sin\theta+\Delta) & -(2\chi\cos\theta+\frac{\gamma} 
{2})%
\end{bmatrix}
.  \label{11}
\end{equation}
The above quadrature operators are defined as, $\delta X=(\delta\mathcal{O} 
^\dag +\delta \mathcal{O})/\sqrt{2}$ and $\delta Y=i(\delta\mathcal{O} 
^\dag-\delta \mathcal{O})/\sqrt{2}$, with $X\equiv(x,I_{i=1,2})$ and $Y\equiv
(p,\varphi_{i=1,2}) $. Similarly, the noises quadratures are given by $\delta 
X^{in}=( \delta \mathcal{O}^{in\dag }+\delta\mathcal{O}^{in})/ \sqrt{2}$ and 
$\delta Y^{in}=i(\delta \mathcal{O}^{in\dag}-\delta \mathcal{O}^{in})/ 
\sqrt{2}$, with $X^{in}\equiv (I_{x}^{in},I_{i=1,2}^{in})$ and $Y^{in}\equiv 
(I_p^{in},\varphi _{i=1,2}^{in})$. The noise operators $\beta^{in}, 
\alpha_1^{in}$, and $\alpha_2^{in}$ have zero mean, and are characterized by 
the following auto correlation functions \cite{[9],[20]}:
\begin{equation}
\begin{cases}
\langle\delta s^{in}(t)\delta s^{in\dag }(t')\rangle=(n_\sigma+1) \delta 
(t-t'), 
\\ 
\langle\delta s^{in\dag}(t)\delta s^{in}(t')\rangle=n_\sigma\delta (t-t'),
\end{cases}
 \label{12}
\end{equation}
with $\delta s^{in}\equiv (\beta ^{in},\alpha_1^{in},\alpha_2^{in}$) and 
$n_\sigma\equiv(n_{th},n_a)$, where $n_{th}=\left[\exp\left(\frac{\hbar\omega_m} 
{k_BT}\right)-1\right] ^{-1}$ and $n_a=\langle\alpha_{1,2}^{in\dag} 
\alpha_{1,2}^{in}\rangle$.

When the system is stable, one gets the following equation for the
steady-state covariance matrix (CM) \cite{[7],[8]}:
\begin{equation}
AV+VA^T=-D.  \label{13}
\end{equation}
Here, the CM is a $6\times 6$ matrix and the elements of the diffusion matrix 
$D$ are defined by 
\begin{equation}
\delta (t-t')D_{i,j}=\frac{1}{2}\langle n_i(t)n_j^\dag(t')+n_j^{\dag}(t) 
n_i(t') \rangle.  
\end{equation} 
Using Eq. (\ref{12}), one obtains
\begin{widetext}
\begin{equation}
D=diag\left[ \frac{\gamma_m}{2}(2n_{th}+1),\frac{\gamma_m}{2}(2n_{th}+1),
\frac{\kappa}{2}(2n_a+1), \frac{\kappa}{2}(2n_a+1),\frac{\gamma }{2}%
(2n_a+1) ,\frac{\gamma}{2}(2n_a+1) \right].
\end{equation}
\end{widetext}
In order to evaluate the entanglement between two subsystems (bipartite
entanglement), the CM should be rewritten as \cite{[8]}
\begin{equation}
V=\begin{pmatrix}
V_\beta & V_{\beta,\alpha_1} & V_{\beta,\alpha_2} \\ 
V_{\beta,\alpha_1}^T & V_{\alpha_1} & V_{\alpha_2,\alpha_1} \\ 
V_{\beta,\alpha_2}^T & V_{\alpha_2,\alpha_1} & V_{\alpha_2}%
\end{pmatrix}
\label{14}
\end{equation}
where each block represents a $2\times 2$ matrix. The blocks on the diagonal 
indicate the variance within each subsystem, while the off-diagonal blocks 
indicate the covariance across different subsystems, i.e., the correlations 
between two components that describe their entanglement property. To compute 
the pairwise entanglement, we reduce the covariance matrix $V$ to a $4\times 4$ 
submatrix $V_S$,
\begin{equation}
V_S=\begin{pmatrix}
V_{k\equiv (\beta,\alpha_1,\alpha_2)} & V_{k,\ell} \\ 
V_{k,\ell}^T & V_{\ell\equiv (\beta,\alpha_1,\alpha_2) }%
\end{pmatrix}
,  \label{15}
\end{equation}
depending on which subsystems we are interested in.

The logarithmic negativity is then defined as \cite{[8]}
\begin{equation}
E_N=\max [0,-\ln 2\eta],  \label{16}
\end{equation}
where 
\begin{equation}
\eta =\sqrt{\frac{\sum-\sqrt{\sum^2-4\det V_S}}{2}},  \label{17}
\end{equation}
is the lowest symplectic eigenvalue of the partial transpose of the
submatrix $V_S$ with 
\begin{equation}
\sum =\det V_k+\det V_\ell-2\det V_{k,\ell}.                               
\end{equation} 

\begin{figure*}[tbh]
\centering
\par
\begin{center}
\resizebox{0.38\textwidth}{!}{
\includegraphics{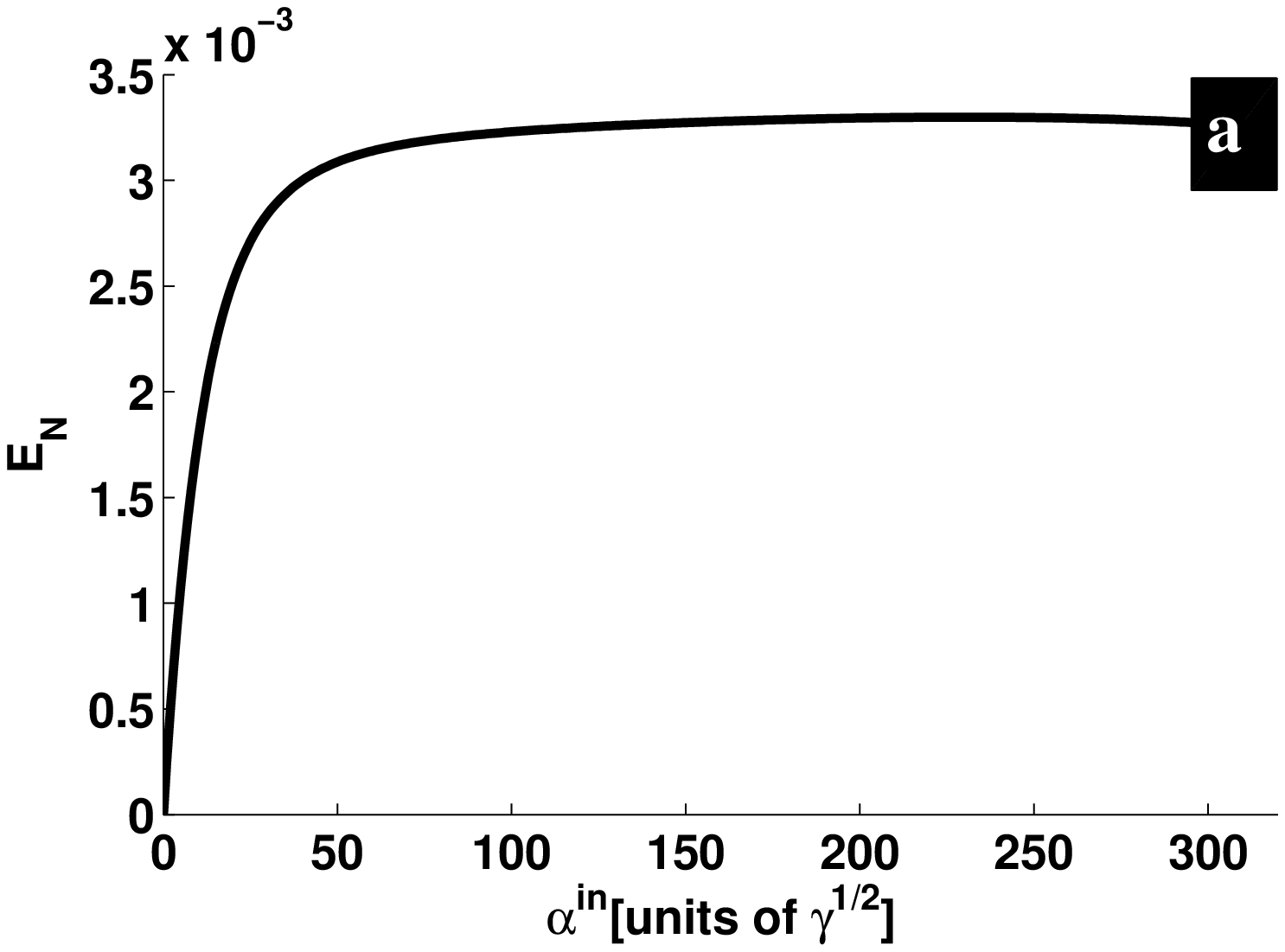}} 
\resizebox{0.38\textwidth}{!}{
\includegraphics{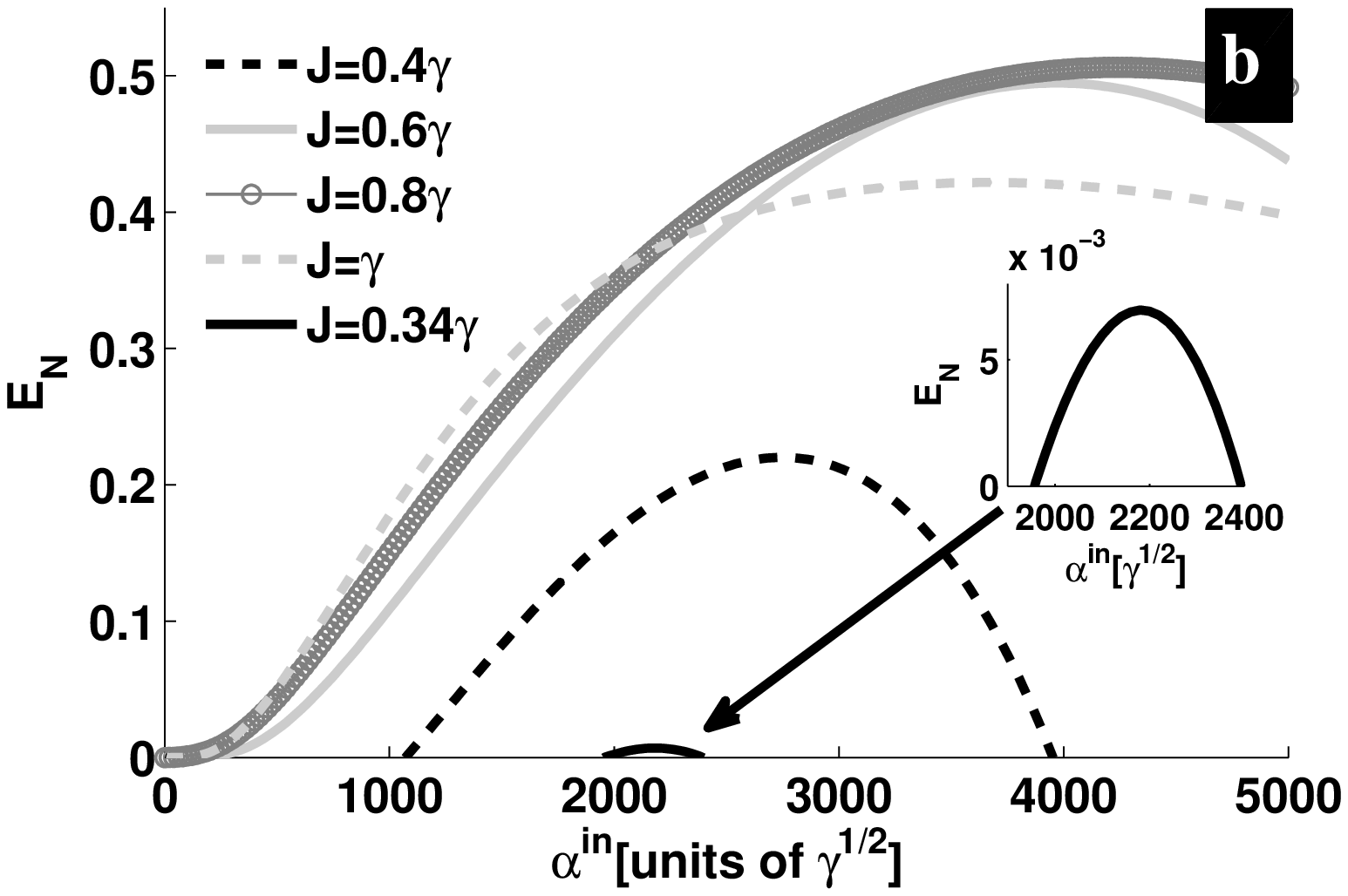}} 
\resizebox{0.38\textwidth}{!}{
\includegraphics{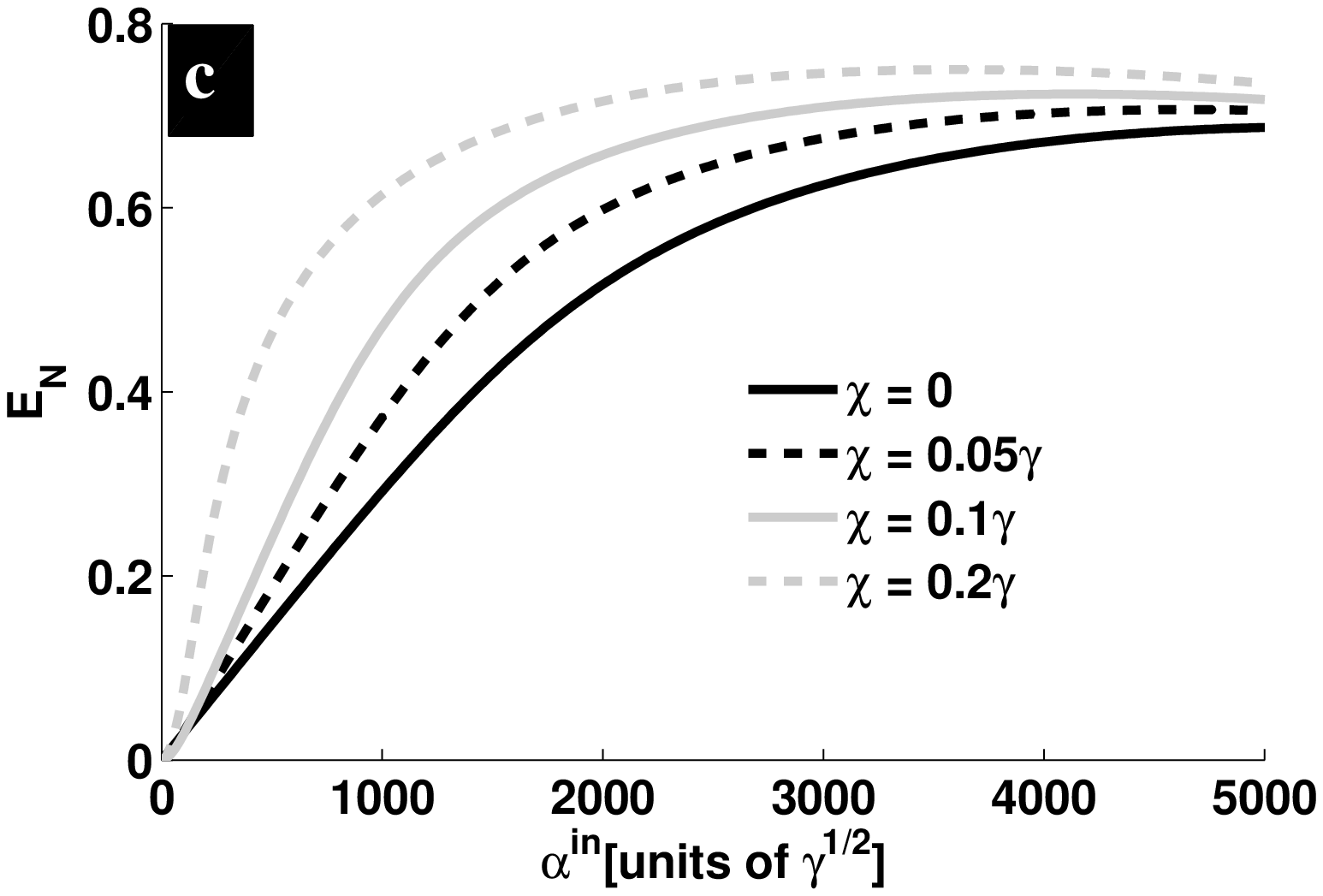}} 
\resizebox{0.38\textwidth}{!}{
\includegraphics{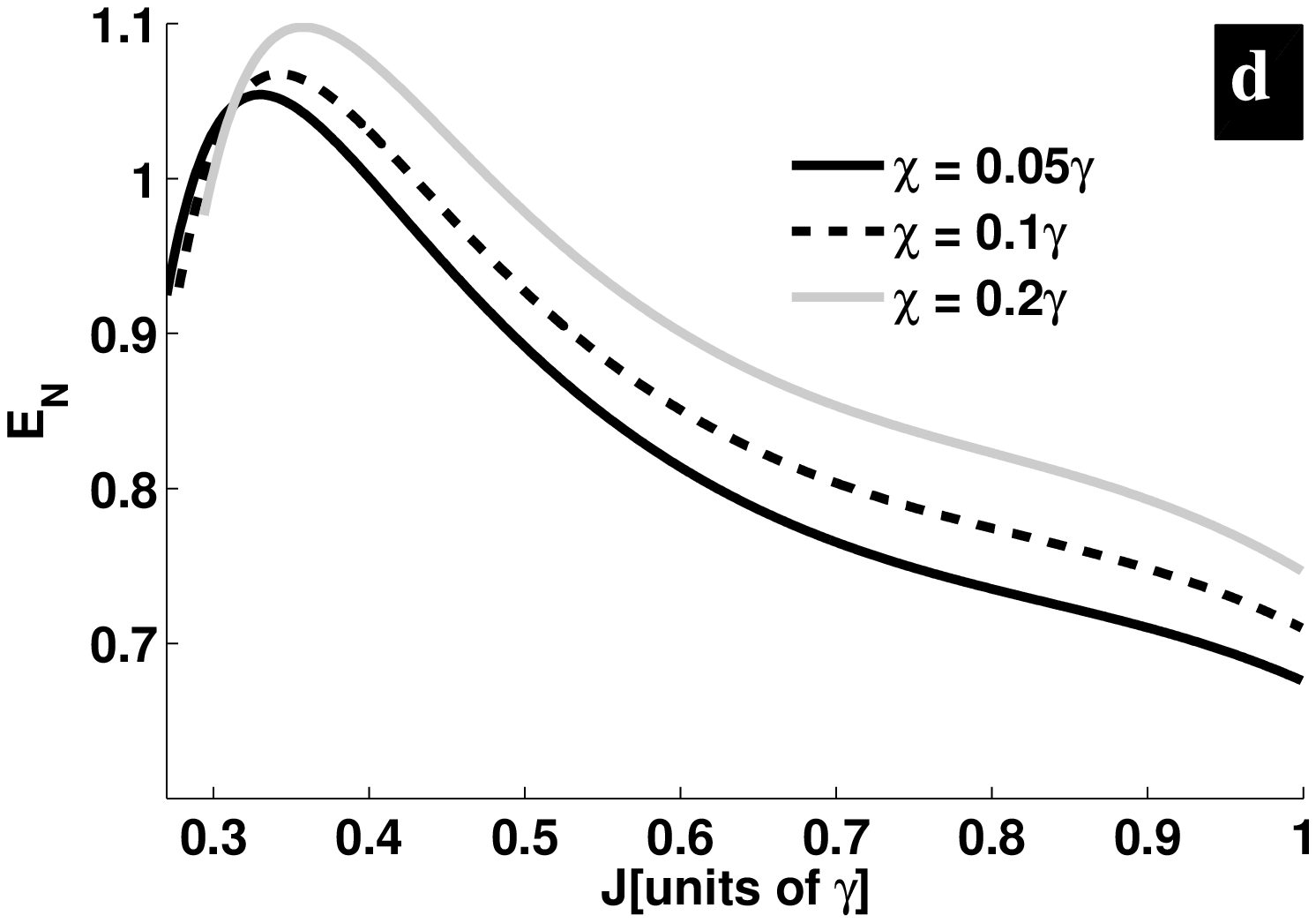}}
\end{center}
\caption{(a) Logarithmic negativity $E_N$ vs $\alpha^{in}$ in the conventional 
COM. (b),(c) The entanglement $E_N$ vs $\alpha^{in}$ between the gain cavity and 
the mechanical modes. (b) $\chi=0$, $\kappa=0.1\gamma$ and different values of 
$J$. (c) $E_N$ vs $\alpha^{in}$, for $\kappa=10^{-5}\gamma$, $J=\gamma$ and 
different values of $\chi$. (d) $E_N$ vs $J$ for $\alpha^{in}=3\times10^3 
\sqrt{\gamma}$, $\kappa=10^{-5}\gamma$ and different values of $\chi$. These 
curves are plotted in the absence of noises ($n_{th}=n_a=0$).}
\label{fig:Fig3}
\end{figure*}

One can now characterize the entanglement through Eq. (\ref{16}). Figure \ref%
{fig:Fig3}(a) shows $E_N$ versus the driving $\alpha^{in}$ in the single lossy 
cavity. One remarks that, the mechanical resonator mode $\beta$ and the optical 
cavity mode $\alpha_2$ are entangled ($E_N>0$). We note that this entanglement 
is weak and is limited by stability conditions that put constraints on the 
magnitude of the couplings $G_{1,2}$  [see black curve in Fig. 
\ref{fig:Fig2}(a)]. This result agrees well with what is predicted in single 
lossy cavity optomechanics \cite{[27]}-\cite{[29]}. In order to enhance 
entanglement, we consider the gain and loss COM system since it improves 
the couplings [see gray curves in Figs. \ref{fig:Fig2}(a) and 
\ref{fig:Fig2}(b)]. As our interest is on distant entanglement, we expect it to 
be enhanced in light of Fig. \ref{fig:Fig2}(b). Figure \ref{fig:Fig3}(b) shows 
entanglement between the gain cavity and the mechanical modes versus 
$\alpha^{in}$. We remark the improvement of the entanglement compared to what 
is generated in the conventional COM [see Fig. \ref{fig:Fig3}(a]). This 
enhancement happens in the unbroken-$\mathcal{PT}$-symmetry regime. Indeed, the 
EP corresponds to $J=0.275\gamma$ and there is no entanglement there (not 
represented). Indeed, the entanglement starts at $J=0.34\gamma$, which is 
beyond the EP [see the black curve in Fig. \ref{fig:Fig3}(b)].  One also remarks 
that, moving from the broken to the unbroken-$\mathcal{PT}$-symmetry regimes, 
the entanglement is enhanced even for weak driving strength [compare black 
curves to the others curves in Fig. \ref{fig:Fig3}(b)]. 

Similar distant entanglement has been investigated in \cite{[41],[42]} by 
coupling two lossy cavities, both supporting a mechanical resonator. Because 
of the absence of $\mathcal{PT}$-symmetry, the generated entanglement as 
relatively weak compared to what is obtained here. Indeed, adiabatic 
approximation was used in \cite{[41]} and both cavities were driven by 
blue-detuned lasers. The resulting amount of entanglement was weak compared to 
what is generated in \cite{[42]}.  However, by driving the cavities by blue- 
and red-detuned lasers, respectively, it is shown that, the amount of generated 
entanglement increases \cite{[42]}. In this work, we point out that 
$\mathcal{PT}$-symmetry has boosted the generation of entanglement in our 
system.

The effect of the PA on the entanglement is shown in Fig. \ref{fig:Fig3}(c). As 
the gain $\chi$ of the PA increases, the entanglement enhances, but mostly for 
relatively weak driving strength $\alpha^{in}$. With the help of the stability 
studied in Sec. \ref{sec:Stab}, the entanglement can be further enhanced as 
shown in Fig. \ref{fig:Fig3}(d). 

\begin{figure*}[tbh]
\centering
\resizebox{0.45\textwidth}{!}{
\includegraphics{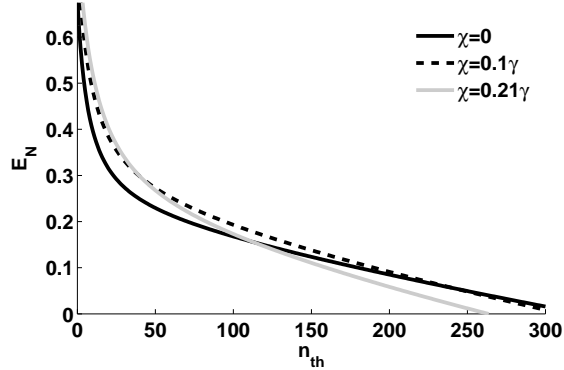}} 
\caption{Logarithmic negativity $E_N$ versus thermal noise $n_{th}$, 
for $\alpha^{in}=3\times10^3\sqrt{\gamma}$, $\kappa=10^{-5}\gamma$, 
$J=0.8\gamma$, and different values of $\chi$.}
\label{fig:Fig4}
\end{figure*}

It is also important to address the effect of noises on the studied 
entanglement. Such concern addresses the robustness of this kind of entanglement 
against decoherence. This is investigated in Fig. \ref{fig:Fig4}, where 
entanglement is plotted versus thermal noise for $n_a=10^{-3}$. One remarks that 
the optical mode inside the gain cavity and the mechanical mode are still 
entangled for thermal noise up to $n_{th} = 300$. However, the weakness of the 
entanglement against thermal noise can be pointed out in the presence of the PA 
(compare gray and black curves in Fig. \ref{fig:Fig4}). It is shown that the 
entanglement is enhanced in gain and loss COM compared to the conventional case. 
Furthermore, this entanglement is improved in the presence of the PA as already 
stated in Refs. \onlinecite{[14]} and \onlinecite{[14a]}. The robustness of 
this entanglement against decoherence is highlighted. 

It is noteworthy that the combined effects of PA and the gain cavity lead to 
instabilities. In order to avoid these instabilities and to investigate 
the effect of $\chi$, we considered the far imbalanced cavities case by 
choosing a small value of $\kappa$ [see Figs. \ref{fig:Fig3} (c), (d) and 
\ref{fig:Fig4}]. 

\begin{figure*}[tbh]
\centering
\par
\begin{center}
\resizebox{0.38\textwidth}{!}{
\includegraphics{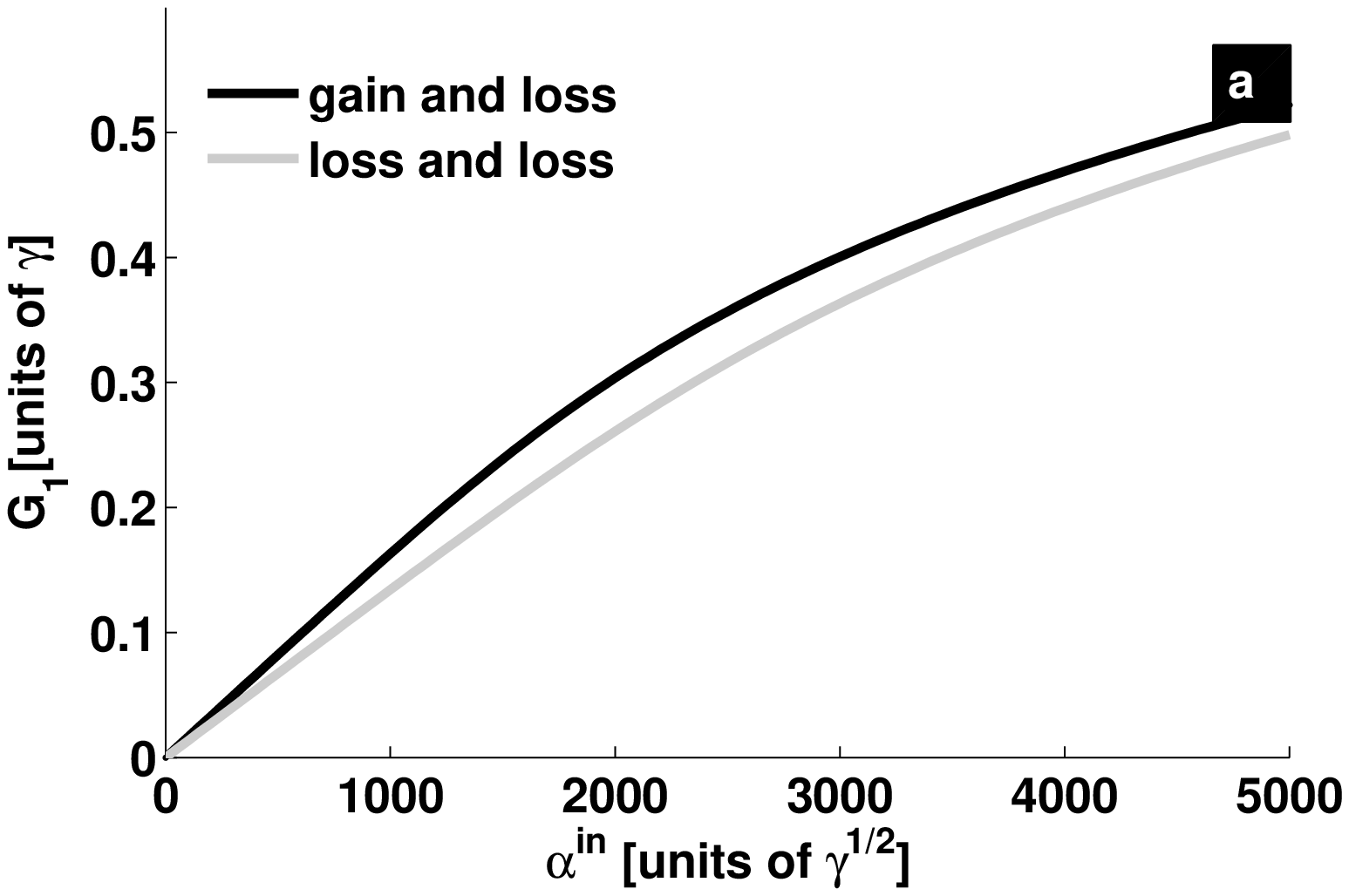}} 
\resizebox{0.38\textwidth}{!}{
\includegraphics{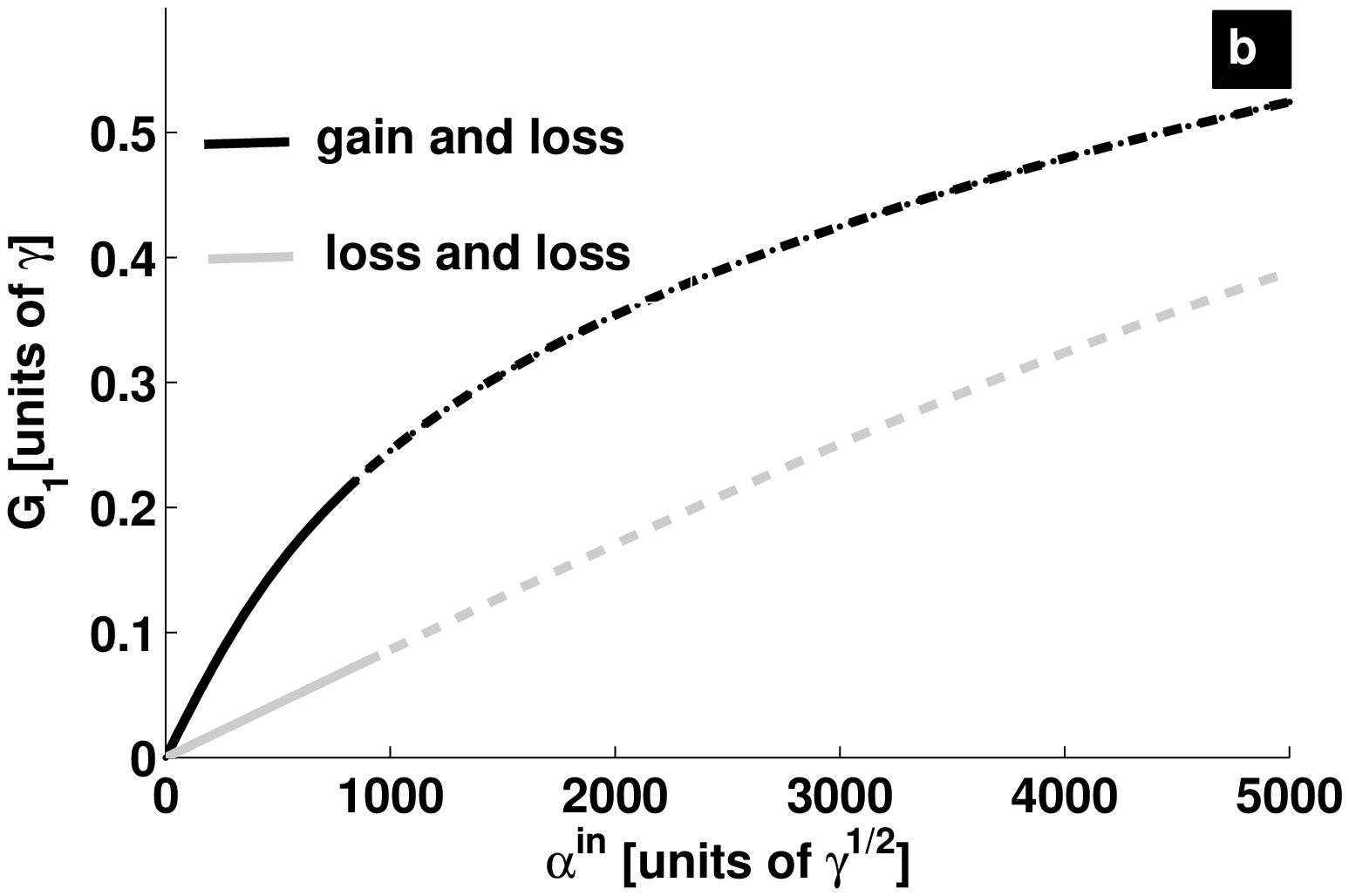}} 
\resizebox{0.38\textwidth}{!}{
\includegraphics{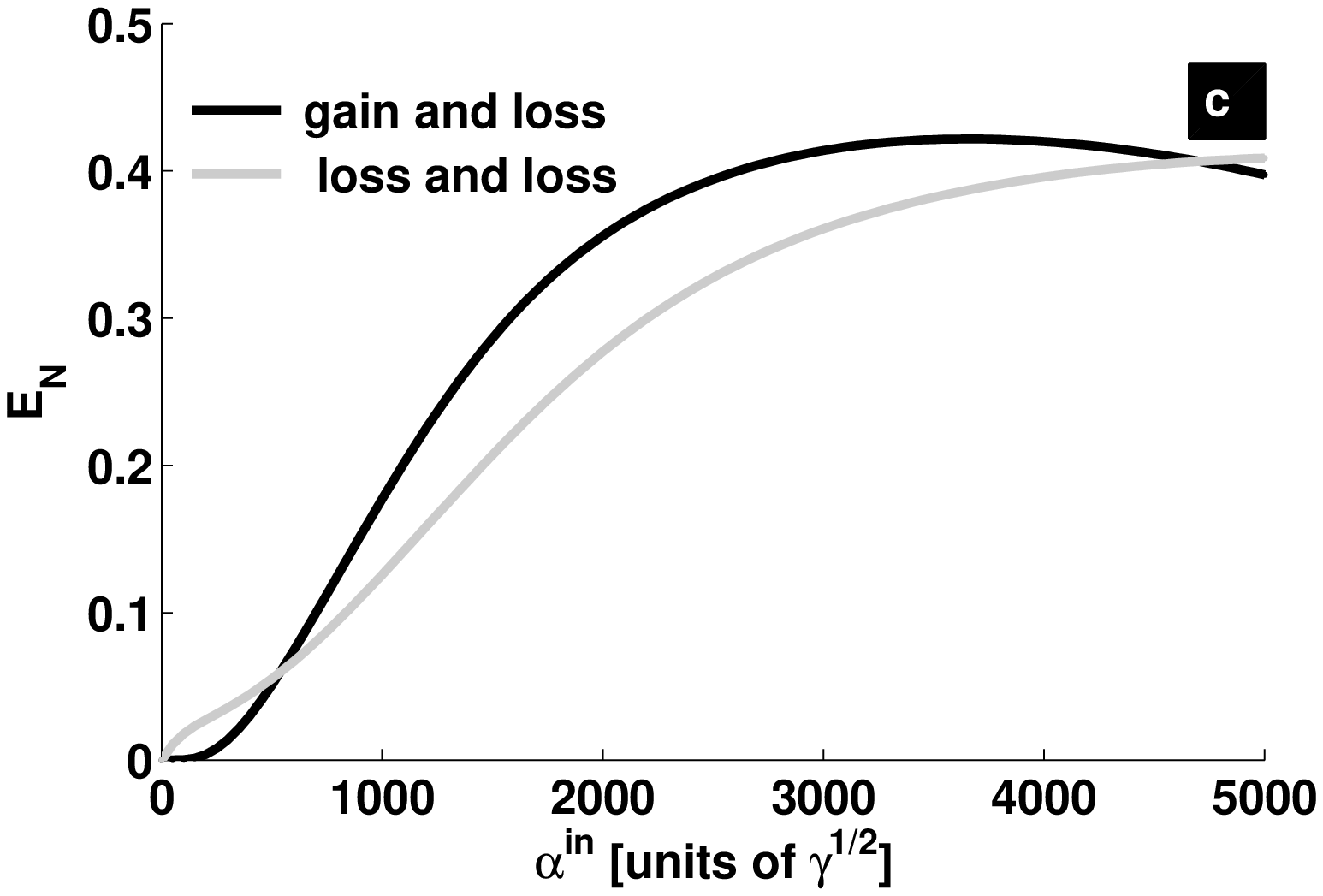}} 
\resizebox{0.38\textwidth}{!}{
\includegraphics{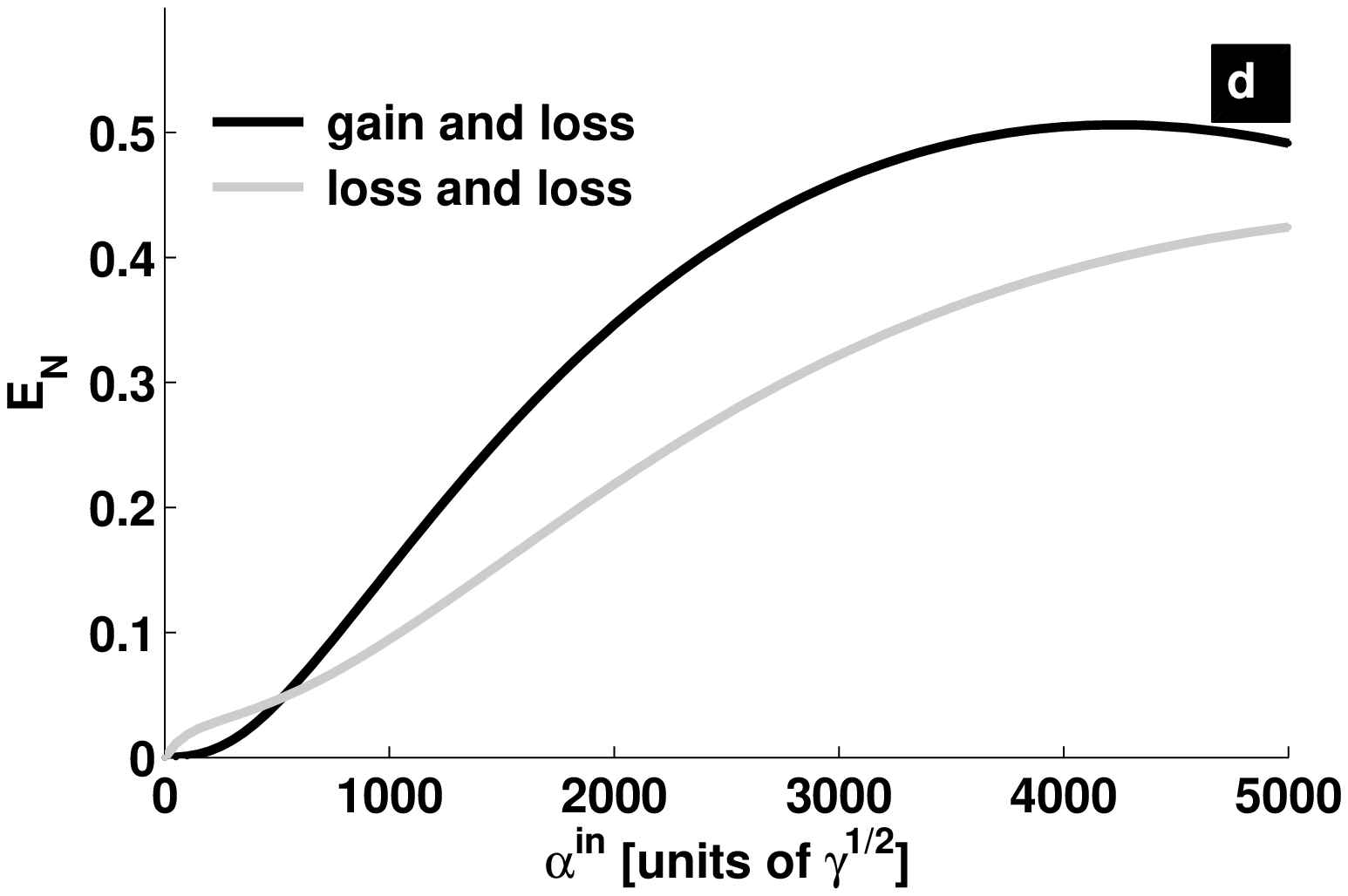}}
\end{center}
\caption{(a),(b) The effective optomechanical coupling $G_1$ versus 
the driving $\alpha^{in}$ for $J=\gamma$, with (a)$\kappa=0.1\gamma$ and (b)
$\kappa =0.8\gamma$. (c),(d) The distant entanglement vs the driving 
$\alpha^{in}$ for $\kappa =0.1\gamma$ with (c) $J=\gamma$ and (d)
$J=0.8\gamma$. Black lines are for the gain and loss coupled cavities 
while the blue lines are for the loss and loss coupled cavities. Full lines are 
stable and the dashed lines are unstable.}
\label{fig:Fig6}
\end{figure*}

In order to show enhancement of entanglement in the $\mathcal{PT}$-symmetry 
system compared to what is generated in a lossy coupled COM, we address a 
comparative study with lossy coupled COM \cite{[41],[42]}. By choosing a 
negative value of $\kappa$ ($\kappa<0$), our coupled COM presented in (Fig. 
\ref{fig:Figa}) becomes a lossy coupled COM \cite{[34],[41],[42]}. The amount 
of distant entanglement is captured by the logarithmic negativity as described 
before. We first compare the couplings resulting in both configurations, and 
then conclude regarding the induced entanglement. Figures. \ref{fig:Fig6}(a) 
and \ref{fig:Fig6}(b) compare the effective coupling $G_1$ for $|\kappa|=0.1 
\gamma$ and $|\kappa|=0.8 \gamma$, respectively. The full lines are stable 
while the dashed ones are unstable. In both cases, the coupling strength is 
improved in the gain and loss COM and this is an indication that entanglement 
will be enhanced accordingly as well. Remarkably, the couplings get closer as 
$|\kappa|$ decreases. Indeed, the couplings are closer for $|\kappa|=0.1 \gamma$ 
[Fig. \ref{fig:Fig6}(a)] than for $|\kappa|=0.8 \gamma$ [Fig. 
\ref{fig:Fig6}(b)]. Furthermore, we have checked that for weak value of 
$\kappa$ ($|\kappa|\approx5\times 10^{-2} \gamma$), there is no difference (at 
least on the coupling strength) between the $\mathcal{PT}$-symmetry COM case and 
the lossy coupled cavities case.  Figures \ref{fig:Fig6}(c) and 
\ref{fig:Fig6}(d)  show distant entanglement generated in both configurations 
for $|\kappa|=0.1 \gamma$ with $J= \gamma$ and $J=0.8 \gamma$ respectively. 
The entanglement captured in Fig. \ref{fig:Fig6}(c) corresponds to the coupling 
shown in Fig. \ref{fig:Fig6}(a). As expected, the generated entanglement in the 
 $\mathcal{PT}$-symmetry configuration is slightly enhanced compared to what is 
obtained in the lossy coupled COM case [see Fig. \ref{fig:Fig6}(c)]. Let us 
keep in mind that, weak coupling rate $J$ can be related to a large distance 
between cavities while strong coupling rate $J$ means a tiny separation between 
them. In this sense, one deduces from  Fig. \ref{fig:Fig3}(d) that, as the 
separation between the cavities increases, the distant entanglement decreases. 
As $J=0.8 \gamma$ in Fig. \ref{fig:Fig6}(d), we deduce that the 
$\mathcal{PT}$-symmetry COM is the best configuration to enhance distant 
entanglement. Indeed, by decreasing the coupling rate $J$ (increasing the 
separation) one gets a net enhancement of distant entanglement in the 
$\mathcal{PT}$-symmetry case compared to what is generated in the lossy coupled 
COM (Fig. \ref{fig:Fig6}d). Thus, $\mathcal{PT}$-symmetry COM is a good 
candidate to enhance distant entanglement. 

\section{Conclusion}

\label{sec:Concl}
We have studied a system of coupled active and passive resonators with the 
focus on steady states stability analysis and the possible generation of distant 
CV entanglement. We have shown through linear stability analysis that, the 
steady-state solutions are generally unstable in the 
broken-$\mathcal{PT}$-symmetry regime and are more stable in the 
unbroken-$\mathcal{PT}$-symmetry phase. The general statement follows that, 
the system is more stable (unstable) for large (small) tunneling coupling rate 
$J$. Conversely, the system is stable (unstable) for small (large) gain-loss 
parameter. We found that the stable solutions correspond to relatively large 
driving strength, compared to the single loss cavity case. Consequently, this 
increases the optomechanical coupling between the mechanical mode and the 
optical field inside the gain cavity. It results in an enhancement of 
steady-state CV entanglement between these modes. It also appears from a 
comparative study with lossy coupled COM that, for weak values of $\kappa$, 
there is no difference between the two cases, regarding the coupling and the 
entanglement; and for large $\kappa$, there is a net enhancement of both 
coupling and entanglement in the gain-loss case compared to what is obtained in 
the loss-loss cavities case. For more entanglement generation, this work 
suggests exploitation the presence of a squeezing element inside the 
active-passive COM. Such nonclassical states can represent an ideal playground 
for investigating and comparing decoherence theories and modifications of 
quantum theory at the macroscopic level.

\begin{acknowledgments}
The authors would like to thank the referees for valuable comments and 
suggestions that have improved the paper.
\end{acknowledgments}

\end{document}